\documentclass[prl,twocolumn,showpacs,preprintnumbers]{revtex4-1}
\usepackage{graphicx}
\usepackage{subfigure}
\usepackage{amsfonts}
\usepackage{amssymb}
\usepackage{amsmath}
\usepackage{bm}

\newcommand{\brho}{\mbox{\boldmath ${\rho}$}}

\begin{document}

\title{Controlled strong coupling and absence of dark polaritons in microcavities with double quantum wells }

\author{K. Sivalertporn$^{1}$}
\author{E.\,A. Muljarov$^{2}$}
\email[]{egor.muljarov@astro.cf.ac.uk}
\affiliation{ $^1$Department of Physics, Faculty of Science, Ubon Ratchathani University, Ubon Ratchathani 34190, Thailand\\
$^2$School of Physics and Astronomy, Cardiff University, Cardiff CF24 3AA, United Kingdom}
\date{\today}

\begin{abstract}
We demonstrate an efficient switching between strong and weak exciton-photon coupling regimes
in microcavity-embedded asymmetric double quantum wells, controlled by an applied electric field. We show that a fine tuning of the electric field leads to drastic changes in the polariton properties, with the polariton ground state being red-shifted by a few meV and having acquired prominent features of a spatially indirect dipolar exciton.
We study the properties of dipolar exciton polaritons, called dipolaritons, on a microscopic level and show that, unlike recent findings, they are not dark polaritons but, owing to the finite size of the excition, are mixed states with comparable contribution of the cavity photon, bright direct, and long-living indirect exciton modes.

\end{abstract}

\maketitle

Semiconductor microcavities are well known due to their importance in fundamental physics and have a wide range of various applications~\cite{Kavokin07}.
Of particular interest are exciton polaritons created in microcavities in the regime of strong light-matter coupling~\cite{Weisbuch92}. The demonstrated possibilities of polariton condensation~\cite{Dend02,Kasprzak06} and room-temperature lasing~\cite{Christopoulos07,Bajoni08,Christmann08} have made microcavity polaritons the subject of intensive studies. If a semiconductor quantum well (QW) is placed in the antinode position of a resonant electro-magnetic field inside a microcavity, Coulomb bound electron-hole pairs localized in the QW layer can strongly interact with a high-quality cavity mode (CM) and form mixed exciton-photon states called polaritons. Double QWs have attracted much attention in recent years due to formation of long-living spatially indirect excitons (IXs) in such structures in the presence of an applied electric field (EF)~\cite{Butov02}. Intensive studies of IXs in double QWs have resulted in demonstration of their electrostatic~\cite{Voros06,Winbow11} and optical control~\cite{Grosso09} as well as in engineering of the dipolar exciton-exciton interaction~\cite{Rapaport05,Schindler08,Cohen11} necessary for exploring different many-body effects, including an intriguing possibility of Bose-Einstein condensation of excitons~\cite{Butov02,Voros06,Yang06,Timofeev07,High12}.

Very recently, asymmetric double quantum wells (ADQWs) have been embedded in a planar Bragg-mirror microcavity \cite{Christmann10} in order to create a special type of polariton with a static dipole moment enhanced at the resonant tunneling condition~\cite{Christmann11}. Indeed, at the electron tunneling resonance, the asymmetry in the conduction band potential of the ADQW is compensated by an applied EF, resulting in a formation of resonant symmetric and antisymmetric electron states while keeping the hole state asymmetric. The electron and the hole are then bound together to form either a direct or an indirect exciton.
When this structure is embedded into the cavity, the direct exciton (DX) strongly couples to the CM and creates a polariton. The IX itself does not form a polariton, as it has a much smaller oscillator strength, but it is electronically coupled to the DX via the tunneling across the barrier. A resulting mixed state of such a three-level system is a polariton that acquires a large electric dipole moment, typical for IX. This allows one to reinforce and control, both electrically and optically, the polariton-polariton interaction. These new hybrid quasi-particles called {\em dipolaritons} were introduced in Refs.~\onlinecite{Christmann11} and \onlinecite{Cristofolini12} and have already been suggested for observation of superradiant terahertz (THz) emission~\cite{Kyriienko12}, continuous THz lasing~\cite{Kristinsson13}, and polariton bistability~\cite{Coulson13}.

It has also been claimed~\cite{Cristofolini12,Szymanska12} that for zero detuning between IX and CM, the middle dipolariton state with the maximum static dipole moment has a vanishing contribution of the DX and consists of only IX and CM, i.e. is a so-called {\em dark polariton}. Physically, this state is similar to the dark counterpart of the superradiant Dicke state~\cite{Dicke54}, and the argumentation which could support this conclusion is analogous to that used in atomic physics in order to describe a dark polariton in a three-level $\Lambda$-system~\cite{Fleischhauer00,Bariani10} and to demonstrate the electromagnetically induced transparency and slowing of light propagation.

We show here that this picture is not however suitable for realistic systems, such as the microcavity-embedded  QWs studied in Ref.\,\onlinecite{Cristofolini12}. We present a microscopic calculation of the optical spectra of such a system and, by making a detailed analysis of the polariton properties, demonstrate the absence of dark polariton states.  We show in particular that the claimed polariton darkness, or vanishing DX contribution, is an artifact of the basis truncation and neglect of the electron-hole (e-h) relative motion. We further demonstrate that an appreciable dipolariton effect, or strong IX-CM mixing, can be achieved only at the cost of a sufficient contribution of the bright DX component.

\begin{figure}[t]
\begin{center}
\includegraphics[width=0.9\columnwidth]{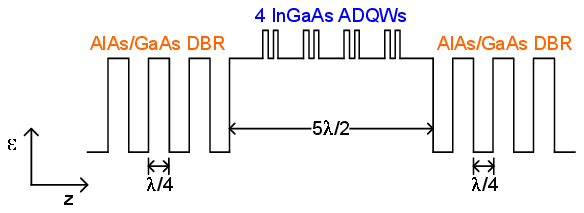}
\caption{Schematic of a microcavity-embedded InGaAs multiple-ADQW structure showing its background dielectric constant.
\label{Structure}}
\end{center}
\end{figure}

The purpose of this Letter is twofold. One is to demonstrate an efficient mechanism of switching on/off the exciton-photon strong coupling regime in cavity-embedded ADQWs and EF control of their polariton properties. We calculate the optical reflection and absorption spectra, analyze them in terms of the exciton-CM coupling and DX-IX crossover~\cite{Sivalertporn12}, and show that the results of our calculation are in quantitative agreement with recent experimental observations~\cite{Christmann10,Christmann11,Cristofolini12}. The other goal is to study dipolaritons on a microscopic level by calculating the full spectrum of exciton states in an ADQW and their coupling to the cavity photon. In this calculation, the polariton brightness and static dipole moment are deduced from the microscopic optical polarization. The relative fractions of the DX, IX, and CM are calculated and analyzed, showing in particular that no dark polariton is observed in such systems.

We concentrate on a microcavity structure used in experiments~\cite{Cristofolini12,Christmann11}, which consists of four InGaAs ADQWs placed at the antinodes of the electromagnetic field inside a 5$\lambda$/2 cavity sandwiched between 17 and 21 pairs of GaAs/AlAs distributed Bragg reflectors (DBRs), see Fig.\,\ref{Structure}. Each ADQW contains two 10-nm In$_{x}$Ga$_{1-x}$As QW layers with the In content of 8\% and 10\% separated by a 4-nm GaAs barrier. To study the light-matter strong coupling in this system, we solve coupled Maxwell's and material equations for the microscopic excitonic polarization $P(z_e,z_h,\brho)$ and the local optical EF ${\cal E}(\omega,k;z)$, using the Green's function approach~\cite{Stahl87,Chuang91,Muljarov02}. Technically, we expand the polarization into the complete set of exciton eigenfunctions:
\begin{equation}
P(z_e,z_h,\brho)= |d_{cv}|^2 \sum_{\nu} \frac{\Psi_{\nu}(z_e,z_h,\brho) X_\nu(\omega,k)}{E_\nu+\hbar^2k^2/2M-\hbar\omega -i\gamma}\,.
\label{Polariz}
\end{equation}
The expansion coefficients are then given by $X_\nu(\omega,k) = \int {\cal E}(\omega,k;z) \Psi_{\nu}(z,z,0) dz$ and are found using the local field ${\cal E}$. The latter in turn satisfies Maxwell's integro-differential equation which includes the macroscopic excitonic polarization $P(z,z,0)$ bringing into the system a nonlocal optical susceptibility~\cite{Nakayama85}. Here $z_e$ ($z_h$) is the electron (hole) coordinate in the growth direction of the ADQW, $\brho$ the coordinate of the e-h in-plane relative motion, $d_{cv}$ the matrix element of the microscopic dipole moment between the valence and conduction bands, $M$ and $k$ the in-plane exciton effective mass and wavevector, and $\omega$ the frequency of the $s$-polarized electro-magnetic field~\cite{footnote2}. To take into account the homogeneous broadening of the excitonic system, a phenomenological damping constant $\gamma$ of all exciton states has been introduced.

\begin{figure}[t]
\begin{center}
\includegraphics[width=0.99\columnwidth,angle=0]{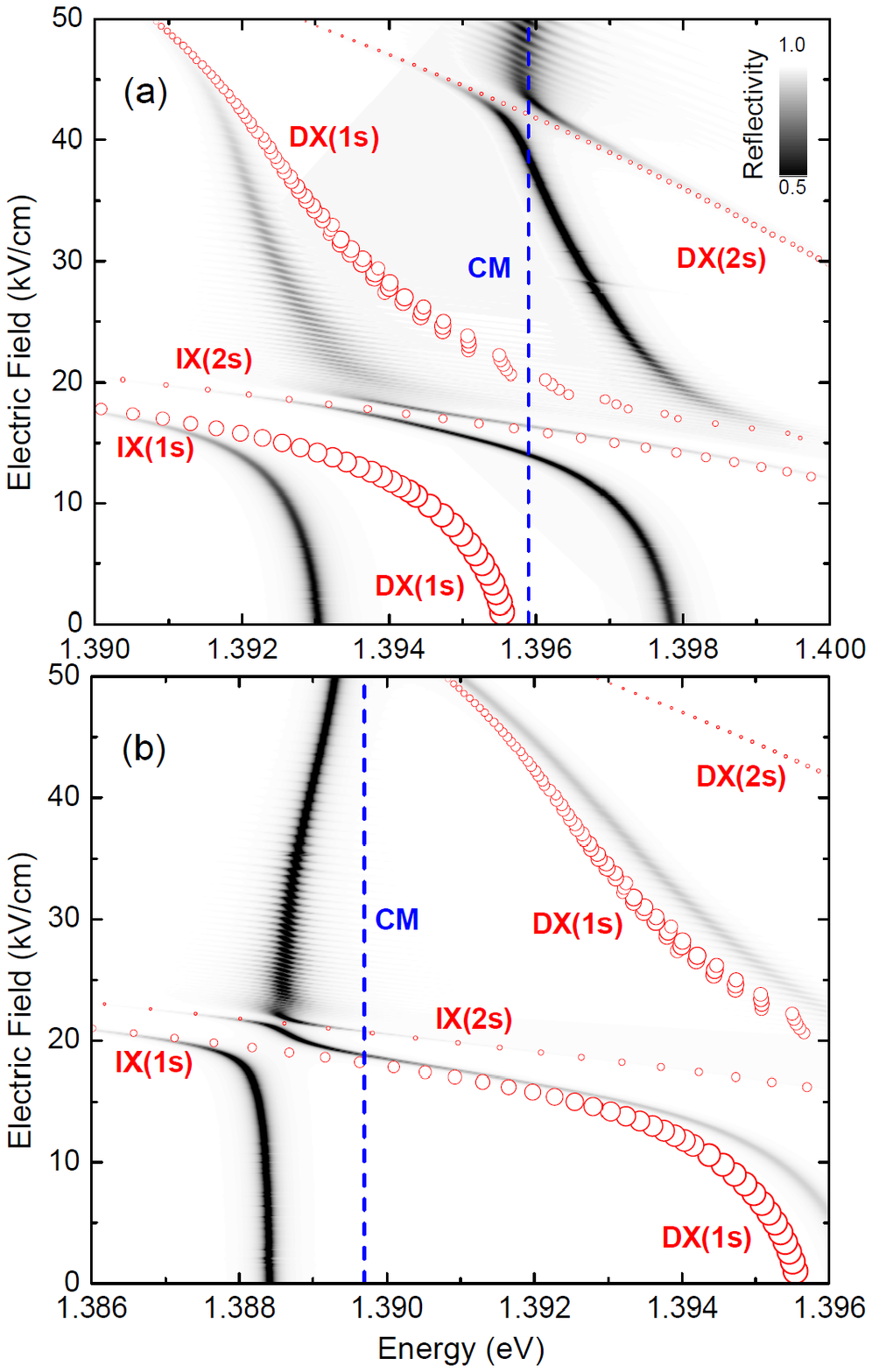}
\caption{Electric field dependent reflectivity spectra of the ADQW structure on a linear grey scale black (0.5) to white (1), for $\gamma=0.1$\,meV and the bare cavity mode at (a) 1.3959\,eV and (b) 1.3897 \,eV (blue dashed lines). The energy positions of four brightest exciton modes are shown by red circles, with the circle area proportional to the exciton oscillator strength. (b) reproduces well the measured reflection shown in Fig.\,3(a) of Ref.~\onlinecite{Christmann11}.
}
\label{Reflectivity}
\end{center}
\end{figure}

In the calculation of the optical polarization Eq.\,(\ref{Polariz}) we keep about 200 exciton states, and for each state $\nu$ calculate its energy $E_\nu$ and the wave function $\Psi_{\nu}(z_e,z_h,\brho)$, by expanding the latter into e-h pair states localized in the ADQW and solving a radial
matrix Schr\"odinger equation for the e-h relative motion in real space~\cite{Sivalertporn12}. Its full solution contains both discrete exciton bound states and unbound states of the excitonic continuum. We discretize this Coulomb continuum by introducing a rigid wall at $|\brho|=R$ with $R=800$\,nm, while keeping the exciton center-of-mass motion untouched. The Maxwell wave equation for the microcavity with ADQWs is solved in a rigorous way, by reducing it to an effective matrix Fredholm problem with a factorizable kernel incorporated into the scattering matrix method~\cite{Ko88,Tikhodeev02}.

Figure~\ref{Reflectivity} shows the reflectivity spectra for different values of the static EF $F$ applied in the growth direction and for two different detunings between the CM (dashed vertical lines) and bare exciton modes (red circles). For small detuning, the four brightest exciton modes of $1s$ and $2s$ type shown in Fig.\,\ref{Reflectivity}(a) by red circles~\cite{footnote1}, are all strongly coupled to the cavity, demonstrating multiple anticrossings. As a result, the spectral positions of the polaritons strongly depend on the EF. In particular, the lowest mode demonstrates a red shift by a few meV, following the exciton GS which in turn experiences a direct-to-indirect  crossover~\cite{Sivalertporn12}. In the case of a larger detuning shown in Fig.\,\ref{Reflectivity}(b), which can be achieved e.g. by changing the cavity width or the angle of light incidence $\theta$, bright exciton states, i.e. those having a considerable contribution of the DX, are still strongly coupled to the CM. However, the polariton properties change abruptly with the EF, due to the $1s$ and $2s$ IX modes crossing the CM -- see in Fig.\,\ref{Reflectivity}(b) the on/off switching of the strong coupling at around $F=20$\,kV/cm.

\begin{figure}[t]
\begin{center}
\includegraphics[width=0.99\columnwidth,angle=0]{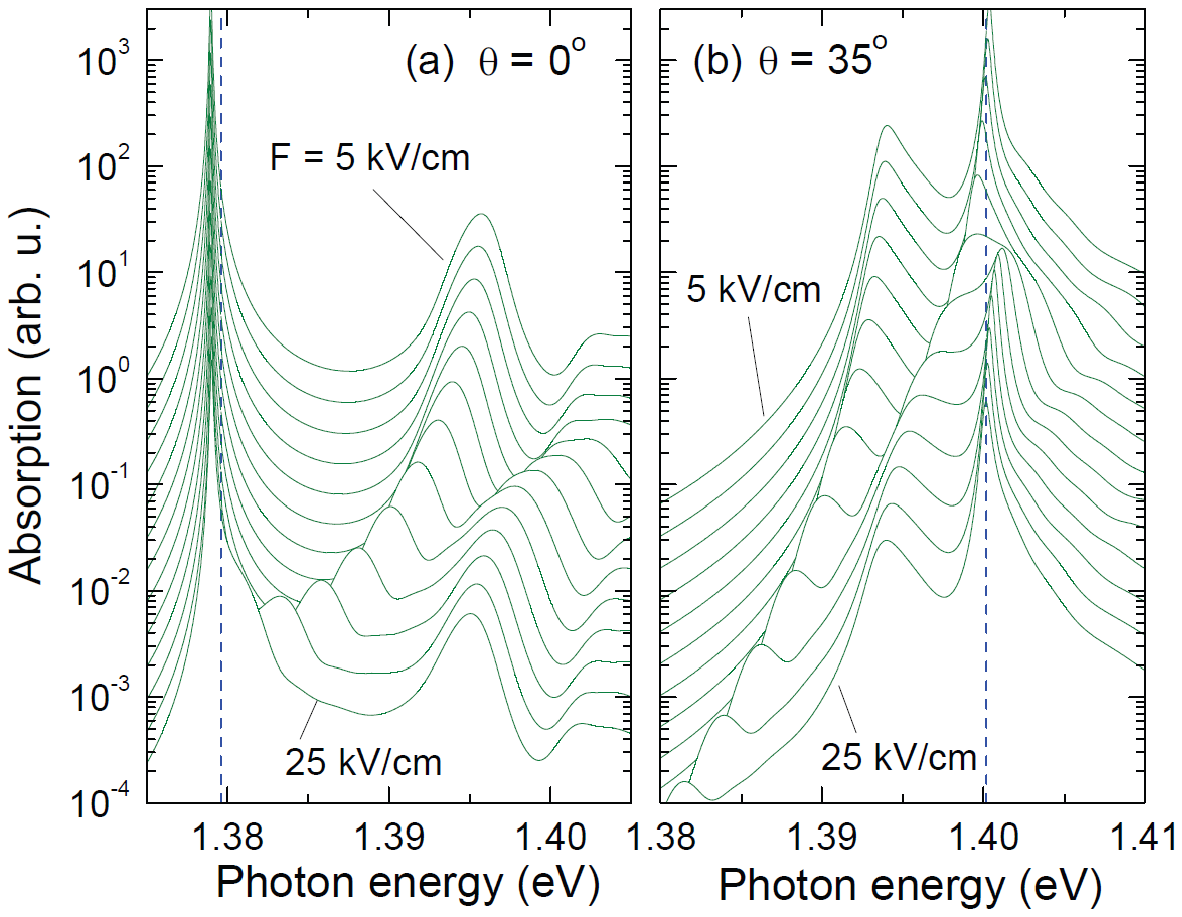}
\caption{Inhomogeneously broadened absorption spectra of the ADQW structure at different EFs, (a) for normal incidence with the bare CM at 1.3796\,eV (blue dashed line) and (b) for $\theta = 35^\circ$ with the bare CM at 1.4002\,eV. Both (a) and (b) reasonably well reproduce the measured photoluminescence spectra shown in Fig.\,2 of Ref.~\onlinecite{Cristofolini12}.
\label{Absorption}}
\end{center}
\end{figure}

The spectrum shown in Fig.\,\ref{Reflectivity}(b) is in good agreement with the measured reflection~\cite{Christmann11}. To reach a quantitative agreement also with the data in Ref.\,\onlinecite{Cristofolini12}, we have calculated the absorption spectra, making a 2\,meV full width at half maximum Gaussian convolution of the excitonic-induced susceptibility, in order to include the actual effect of inhomogeneous broadening of the exciton lines~\cite{Andreani98}. The result is shown in Fig.\,\ref{Absorption} for two different angles of incidence. At normal incidence, the CM is weakly coupled to ADQWs as demonstrated by a narrow peak in the absorption. This regime changes to the strong coupling for non-normal incidence: At $\theta=35^{\circ}$, one can clearly see three wide spectral bands corresponding to the three polariton modes which were observed in Ref.\,\onlinecite{Cristofolini12} and analyzed there in terms of the above mentioned three-level model (TLM). The $\theta = 35^\circ$ spectra were used for the fit of the detuning, for details see Fig.\,S1 of the supplementary material (SM). Apart from the missing temperature-dependent state occupation factor and a rigid shift of all spectra by 30\,meV, a quantitative agreement with the measured photoluminescence~\cite{Cristofolini12} is achieved that demonstrates the quality of our approach and justifies our microscopic analysis of dipolariton states given below.
\begin{figure}[t]
\begin{center}
\includegraphics[width=0.99\columnwidth,angle=0]{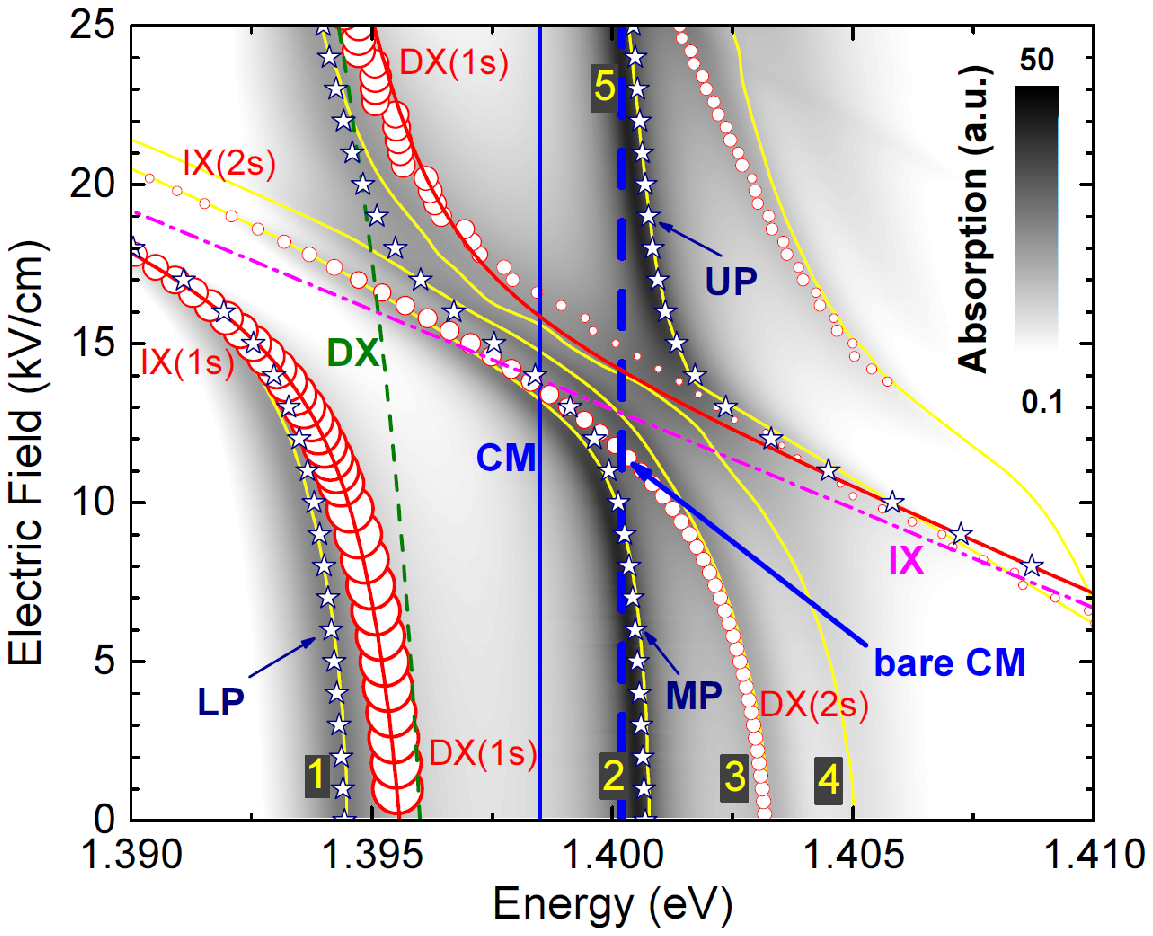}
\caption{Electric field dependent absorption spectrum of the ADQW structure on a logarithmic grey scale from white (0.1) to black (50),
for $\theta=35^\circ$, the bare CM position at 1.4002\,eV (blue dashed line), $\gamma=0.5$\,meV, and no inhomogeneous broadening. Polariton states at the peak positions of the absorption are numbered and indicated by yellow dashed lines. Polariton modes calculated within the TLM~\cite{Cristofolini12} are shown by stars, with the uncoupled CM, DX, and IX modes given by solid blue, dashed green, and magenta lines, respectively, and coupled DX and IX modes by two solid red lines matching well the microscopic calculation (red circles).
\label{Polarstate}}
\end{center}
\end{figure}

To study the dipolaritons we switch off the inhomogeneous broadening in our calculation and use the strong coupling case of $\theta=35^{\circ}$ discussed above. Also, we have increased the damping to $\gamma=0.5$\,meV, in order to smear out any signatures of the artificial discretization of the Coulomb continuum, well resolved in Fig.\,\ref{Reflectivity}.  The result is seen in Fig.\,\ref{Polarstate} as a series of peaks in the absorption spectrum [similar to Fig.\,\ref{Absorption}(b)]-- dark regions emphasized by yellow dashed lines with numbers indicating the energy positions of the polariton states. Unlike the TLM of Ref.\,\onlinecite{Cristofolini12}, the DX and IX in the present approach are neither eigenstates of the system, nor any basis states. Thus, for an adequate comparison with the TLM, we introduce the fractions of DX, IX, and CM in each polariton state by using  polariton brightness ${\cal F}$ and dipole moment ${\cal D}$ defined as
\begin{eqnarray}
{\cal F} &=&{\cal N}^{-1} \left|\int P(z,z,0) dz\right|^2\,,\\
{\cal D} &=& {\cal N}^{-1}\int\!\!\! \int\!\!\! \int \left|P(z_e,z_h,\brho)\right|^2 (z_e-z_h) d\brho\,dz_e\,dz_h\,,\\
{\cal N} &=& \int \!\!\!\int \!\!\!\int \left|P(z_e,z_h,\brho)\right|^2 d\brho\,dz_e\,dz_h\,,
\end{eqnarray}
where ${\cal N}$ is a normalization integral. The relationship between the DX, IX, and CM components ($C_{\rm DX}$, $C_{\rm IX}$, and $C_{\rm CM}$, respectively) is then given by
\begin{equation}
\left({C_{\rm IX}}/{C_{\rm X}}\right)^2\approx \alpha{\cal D}\,,\ \ \ \ \
\left({C_{\rm CM}}/{C_{\rm X}}\right)^2 \approx \beta{\cal F}\,,
\end{equation}
with $C_{{\rm X}}^2 = C_{{\rm DX}}^2 + C_{{\rm IX}}^2$. Indeed, the IX component is accurately determined by the dipole moment, with the darkest exciton corresponding to the maximum e-h separation. Therefore at its maximum value ${\cal D}\approx d^\ast$, so that $\alpha=1/d^\ast$, where $d^\ast$ is the mean e-h separation in the indirect exciton (details of its calculation are given in Sec.\,II of the SM), which for $F=0$ almost coincides with  the center-to-center distance $d$ between QWs in the ADQW. The CM component is determined by the brightness ${\cal F}$ which is proportional to the exciton oscillator strength and consequently to the exciton-CM coupling. Then $\beta$ is evaluated by mapping the coupled Maxwell's and material equations onto the exciton-photon Hamiltonian which leads to $\beta=e^2 |d_{cv}|^2\epsilon_b( \bar{\cal E}/\bar{P})^2/(2\pi E_{\rm CM}d)$, where $E_{\rm CM}$ is the CM energy, $\epsilon_b$ is the background dielectric constant, and $\bar{\cal E}$ and $\bar{P}$ are the mean values of the local field and polarization inside the ADQW: $\bar{P}^2d\approx \int |P(z,z,0)|^2 dz$. Normalized fractions $C^2_{\rm DX}$, $C^2_{\rm IX}$, and $C^2_{\rm CM}$ are then calculated by taking the values of  ${\cal F}$, ${\cal D}$, and ${\cal N}$ at the polariton frequencies.

\begin{figure}[t]
\begin{center}
\includegraphics[width=0.8\columnwidth]{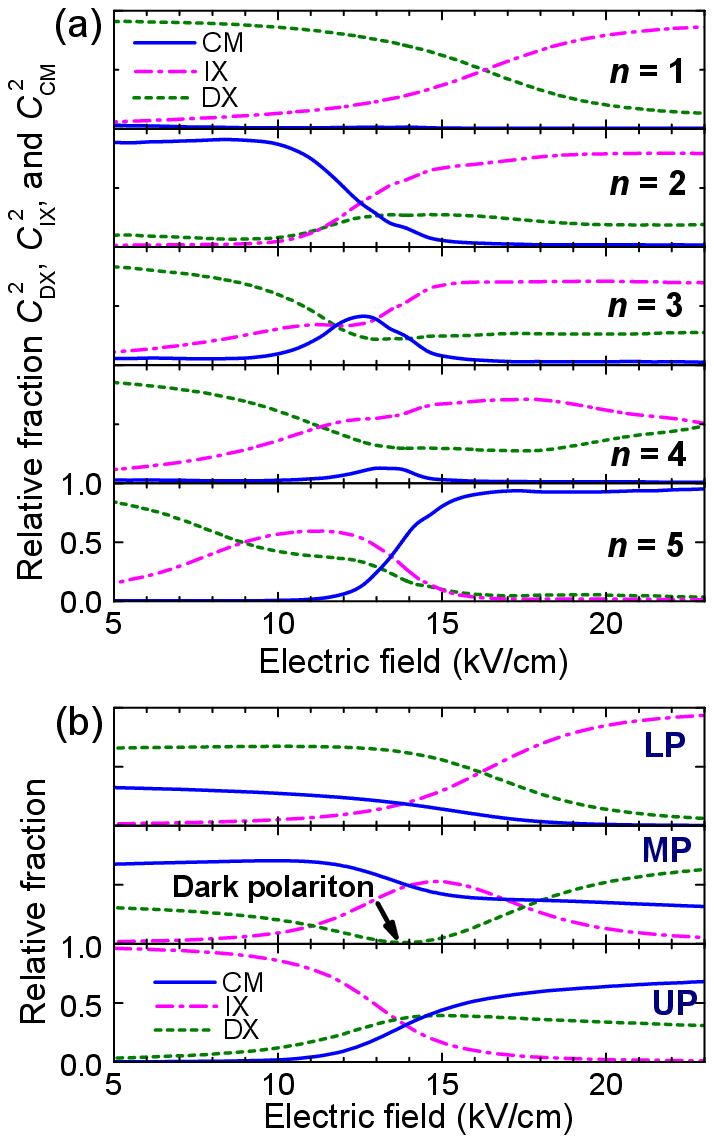}
\caption{(a) Relative fractions of DX, IX and CM in the polariton states $n=1$ to 5, seen as maxima in the absorption spectra in Fig.\,\ref{Polarstate}, as functions of the EF, for $\theta=35^\circ$. (b) The same fractions calculated within the TLM~\cite{Cristofolini12} for LP, MP, and UP branches shown in Fig.\,\ref{Polarstate}. Black arrow in (b) marks position of the dark polariton predicted by the TLM, which not seen however in the full calculation (a).
\label{Fraction}}
\end{center}
\end{figure}

The relative contribution of DX, IX, and CM to the lowest five polariton states is shown in Fig.\,\ref{Fraction}\,(a). The ground state ($n=1$) has a small CM component -- the corresponding exciton state is detuned from the CM -- and demonstrates a DX-IX crossover typical for the exciton GS. The first ($2s$) and higher excited exciton states producing polariton modes $n=2$ to 5 are resonant to the bare CM for some values of the EF (see Fig.\,\ref{Polarstate}) and thus demonstrate a considerable CM contribution. For example, state $n=2$ ($n=5$) follows closely the bare CM (dashed blue line in Fig.\,\ref{Polarstate}) at small (large) values of the EF which results in the dominant contribution of the CM to the polariton, as seen in Fig.\,\ref{Fraction}\,(a). Importantly, their excitonic part consists of IX and DX contributions of similar strength. The maximum dipolariton effect is achieved for states with $n=2,$\,3 and 5 at around $F=12.5$, 12, and 13.5\,kV/cm, respectively, when the contribution of all three components is nearly the same ($\sim1/3$). However, an appreciable static dipole moment of the polariton is achieved only together with an appreciable DX fraction. Varying the detuning can reduce the influence of $2s$ and higher exciton states, resulting in a ``darker'' polariton, but this always happens at the cost of a significant reduction of the exciton fraction, as demonstrated by Figs.\,S4 and S5 of the SM.

Finally,  Fig.\,\ref{Fraction}\,(b) repeats the result of the TLM~\cite{Cristofolini12}, with the relative positions of the DX, IX, and CM and the coupling parameters taken the same as in Ref.\,\onlinecite{Cristofolini12}. The microscopic energies of ground and the second excited exciton states are well reproduced by this DX-IX coupling, compare red lines and circles in Fig.\,\ref{Polarstate}. The lower (LP), middle (MP) and upper polariton (UP) branches of the TLM also match quite well the positions of the most prominent polariton resonances obtained in the full calculation, compare dark blue stars with yellow dashed lines or dark regions in Fig.\,\ref{Polarstate}. However, there is no good correlation between the CM, IX, and DX fractions calculated in both approaches as it is clear from Fig.\,\ref{Fraction}. In particular, there is no dark polariton observed in the microscopic calculation. Furthermore, the bare CM (at 1.4002\,eV), corresponding to the spectral minimum in the microcavity reflection calculated without the ADQW structure, does not match the CM in the TLM fit (at 1.3985\,eV).

We see the main reason for these disagreements in the large number of exciton states taken into account in our approach and in a significant role of the Coulomb interaction affecting the e-h relative motion which is {\em a priori} missing in the TML. Indeed, the usual modeling of cavity polaritons in terms of a matrix Hamiltonian~\cite{Skolnick98} consideres only the lowest ($1s$) exciton states assuming that much smaller oscillator strengths of $2s$ and higher exciton states result in their negligible impact on the polariton. In the present case of dipolaritons, the influence of these higher states becomes more perceptible, as the detuning condition leading to a vanishing DX component cannot be fulfilled for all excitonic states simultaneously. Furthermore, we believe that the concept of dark polaritons is contradictory in itself as the ``darkness'', originating from spatially indirect states, implies a finite exciton size and existence of its internal structure. These features are however fully neglected in the matrix approach. In other words, particles treated as point-like objects have infinite mass of their relative motion. The latter in turn prohibits any tunneling required for creation of dark polaritons.

In conclusion, we have demonstrated, using a microscopic calculation, the existence of hybrid exciton polariton states with a large static dipole moment, called dipolaritons, and have studied their properties. We show that the dipolaritons have comparable contributions of long-living indirect exciton, optically bright direct exciton and photon cavity mode. This is due to the exciton internal structure and in particular to $2s$ and higher excited states which show up as additional unticrossings in optical spectra and spoil the destructive interference responsible for creation of dark polaritons. This clarifies in particular that the dark polaritons claimed in Refs.~\onlinecite{Cristofolini12,Szymanska12} are an artifact of a three-level matrix Hamiltonian used there and further suggests a more general conclusion on the absence of dark polaritons in principle, in any realistic, finite-size system. The reflectivity and absorption spectra predicted here are in quantitative agreement with experimental observations and demonstrate an efficient switching between weak and strong coupling regimes, controlled by the applied electric field, which is promising e.g. for applications as a fast and small electro-optical modulator.
\bigskip

The authors thank W. Langbein, V. D. Kulakovskii, and J. J. Baumberg for discussions.
K.\,S. acknowledges support of the Royal Thai Government.


\begin{thebibliography}{99}
\bibitem{footnote2} We use the same parameters as in Ref.\,\onlinecite{Sivalertporn12}, including $d_{cv}=0.6$\,nm.
\bibitem{footnote1} More exciton states in the same system are shown in Fig.\,9 of Ref.~\onlinecite{Sivalertporn12}.
\end{thebibliography}
\end{document}


\title{Supplementary material: Controlled strong coupling and absence of dark polaritons in microcavities with double quantum wells}
\author{K. Sivalertporn$^{1}$}
\author{E.\,A. Muljarov$^{2}$}
\affiliation{ $^1$Department of Physics, Faculty of Science, Ubon Ratchathani University, Ubon Ratchathani 34190, Thailand\\
$^2$School of Physics and Astronomy, Cardiff University, Cardiff CF24 3AA, United Kingdom}
%
%
%
\maketitle

\section{Calculated absorption vs measured photoluminescence}

Here we compare experimentally measured~[20] photoluminescence (PL) spectra of the microcavity-embedded asymmetric double quantum well (ADQW) structures with inhomogeneously broadened absorption spectra calculated at different values of the detuning which is defined as the energy difference between the cavity mode (CM) and the exciton ground state (GS) energy at zero electric field (EF). The aim of this comparison is to find the optimal value of the detuning at which the experimental spectra are best reproduced by the microscopic calculation.
Apart from the phenomenological broadening $\gamma$, the detuning is the only unknown parameter of the present calculation as both the semiconductor band gap and the CM position are quite sensitive to the temperature of the sample and thus are not determined well. Other parameters of the calculation, which are either structural or material parameters, are all well determined.

\begin{figure}[b]
\includegraphics*[angle=0,width=0.99\columnwidth]{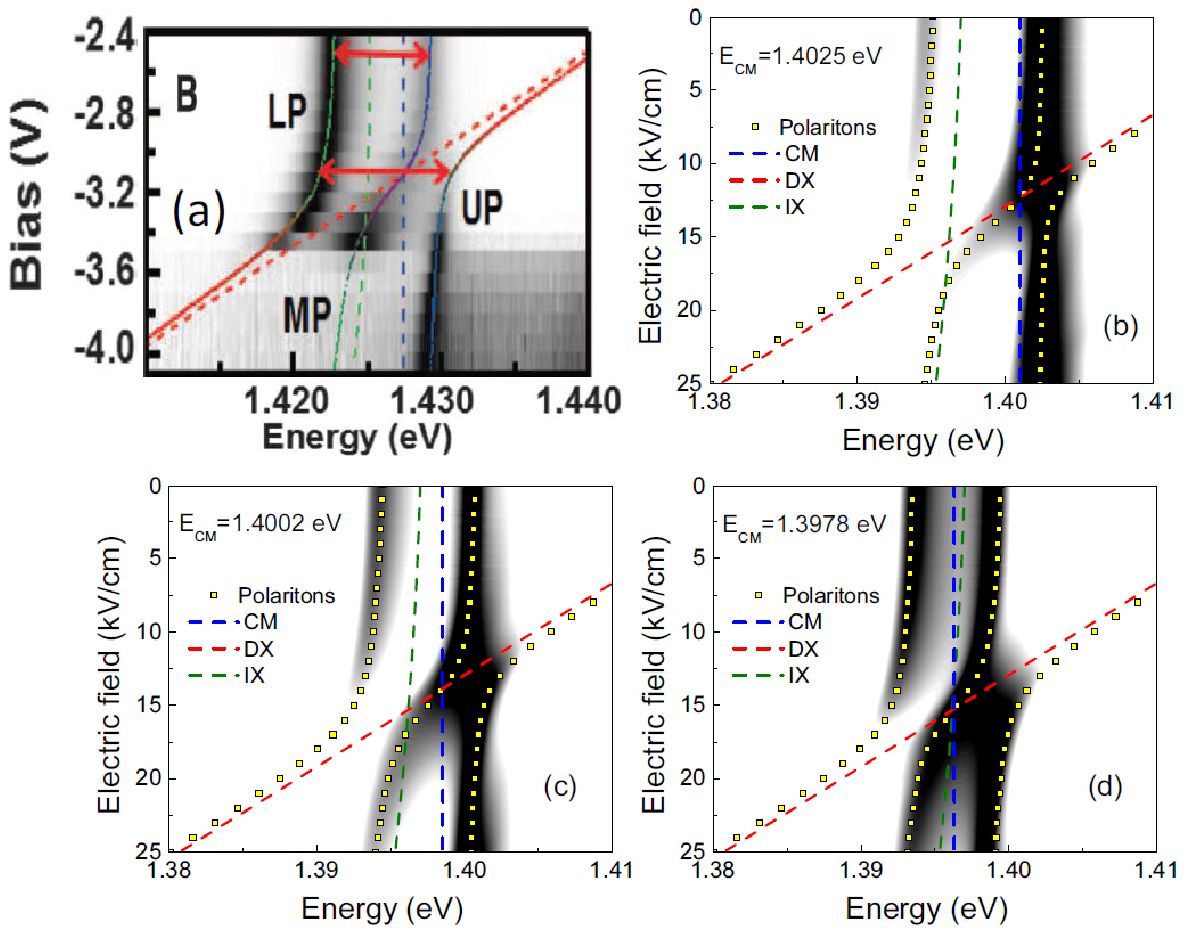}
\caption{(a) Measured PL spectrum of the biased microcavity-embedded ADQW -- the figure is copied from Fig.\,2A of Ref.\,20, in order to make the comparison more explicit. (b)-(d) Absorption spectra showing 2.5 orders of magnitude in log scale (from back to white), calculated for three different detunings with the bare CM positions as given. Fitted DX, IX and  CM states (dashed lines) and polariton states (rectangles) calculated within the three-level model using the coupling parameters $\Omega=J=6$\,meV, the same as in Ref.\,20. The energies of the DX and IX states  in EF are given by the fitted dependence in Eq.\,(\ref{EDXIX}).
}\label{fig:SM1}
\end{figure}

In addition to the adjustment of the detuning, we also analyze here the calculated absorption using the three-level model (TLM) introduced in Ref.~20, by fitting the positions of the maxima in the absorption with the upper (UP), middle (MP), and lower polariton (LP) levels of the TLM.
For comparison, we have taken the measured PL spectrum from Fig.\,2A of Ref.~20, and shown it in Fig.\,\ref{fig:SM1}(a) along with the absorption in Fig.\,\ref{fig:SM1}(b)-(d), calculated for $\theta = 35^\circ$ and three different values of the detuning (the bare CM positions are as given in the figure), with $\gamma=0.5$\,meV homogeneous and 2\,meV half-width-at-half-maximum inhomogeneous broadening, as described in the main text. The rigid shift of the calculated spectra of about 30\,meV as compared to the measured PL lines, seen in Fig.\,\ref{fig:SM1}, is unimportant as removing it would lead to negligible corrections in the spectra. Important is only the detuning which drastically modifies the spectra. We have used a polynomial approximation to describe the EF dependence of the bare direct (DX) and indirect exciton (IX) energies,
\be
E_{\rm DX}(F)=1.397-3.0\cdot 10^{-5} F-1.5\cdot10^{-7}F^2\ \ \ \ {\rm and}\ \ \ \ \
E_{\rm IX}(F)=1.420-1.6\cdot10^{-3} F\,,
\tag{S\theequation}\label{EDXIX}\stepcounter{equation}
\ee
respectively, where the energies are measured in eV and the EF $F$ in kV/cm. We found the above dependencies by fitting the microscopically calculated exciton spectrum [28] with the use of the DX-IX coupling constant $J=6$\,meV, taken the same as in Ref.\,20.

We think the detuning used for the spectrum in Fig.\,\ref{fig:SM1}(b) provides the best fit to the experiment, although this may seem to be not so obvious due to the fact that the PL and the absorption spectra differ by an exponential thermal-distribution factor. Indeed, using the parameters of Fig.\,\ref{fig:SM1}(b) also for the normal incidence (corresponding to the weak coupling regime, as discussed in the main text) reproduces well the measured distance between the two lowest polariton lines, which is not achievable for any other detuning. Nevertheless, we decided to use in the main text the detuning producing the absorption in Fig.\,\ref{fig:SM1}(c), for which the TLM fit obtained almost coincides with that reported in Ref.~20 -- compare the CM, DX, and IX lines in Figs.\,\ref{fig:SM1}(a) and (c). However, for completeness of our analysis, the analogues of Figs.\,4 and 5 calculated with the  detuning used in Fig.\,\ref{fig:SM1}(b) are also given in Sec.\,III below.

\section{Mean electron-hole separation in the indirect exciton}

\begin{figure}[t]
\hskip-0.0cm
\includegraphics*[angle=0,width=0.99\columnwidth]{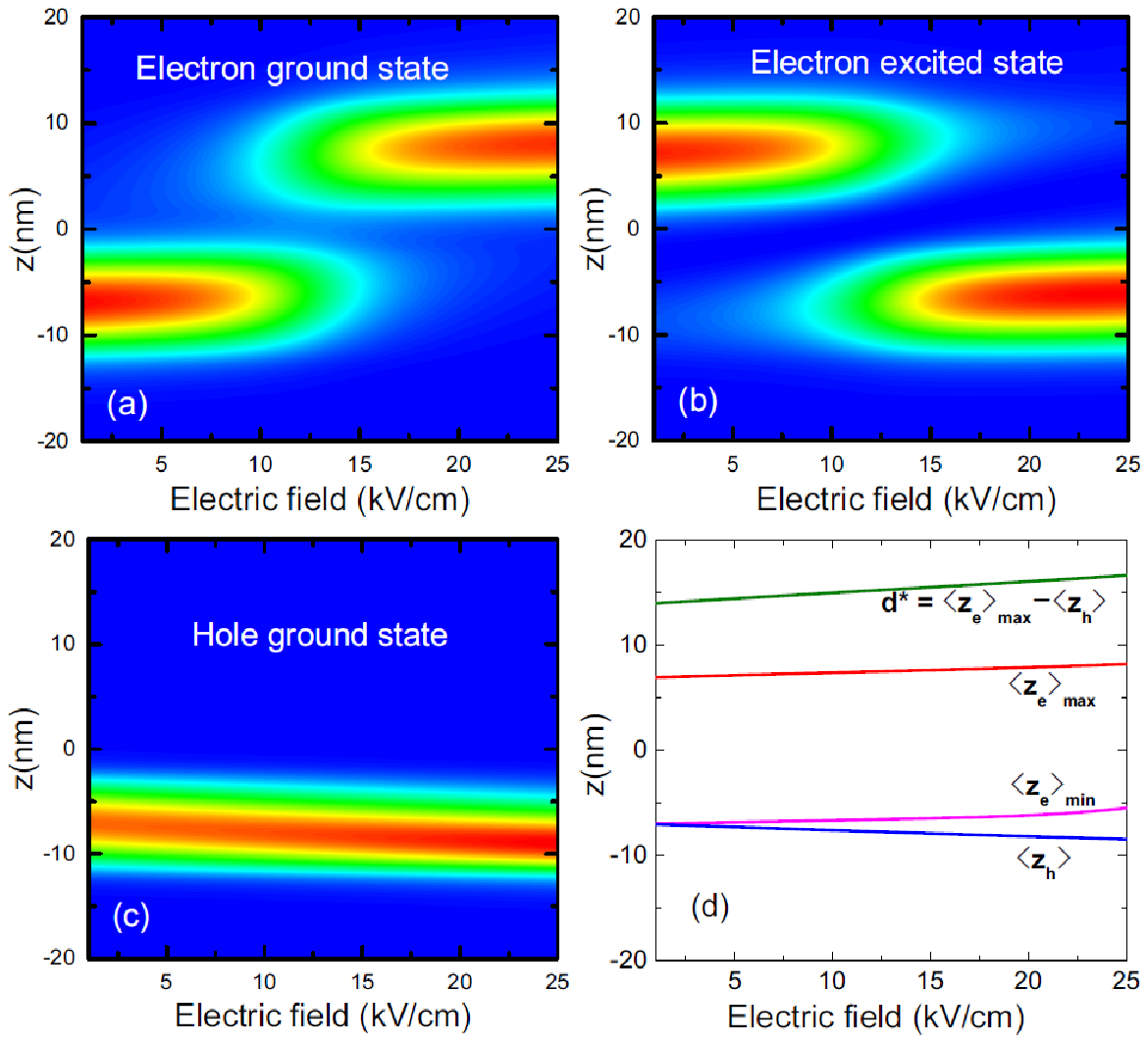}
\caption{The wave function squared of the electron ground (a) and excited state (b) as well as of the hole ground state (c) in the ADQW in an electric field. (d) The hole ground state mean coordinate $\langle z_h\rangle$ and the electron mean coordinate $\langle z_e\rangle_{\rm min}$ and $\langle z_e\rangle_{\rm max}$ in the DX and IX, respectively, as well as the mean e-h separation $d^\ast$ in the IX, as functions of the electric field.
}\label{fig:SM2}
\end{figure}

This section provides details on how the electron-hole (e-h) separation in the IX and its change with the EF is calculated in the present approach. The EF modifies the QW and consequently the electron and hole states localized in it. Figs.\,\ref{fig:SM2}(a)--(c) show plots of the wave function of the lowest electron and hole states versus the EF $F$. For small values of the EF (up to 5-7\,kV/cm), the IX wave function is dominated by the e-h pair state consisting of the hole ground and the electron excited states. Therefore the e-h mean separation $d^\ast$ in the IX in this regime can be found simply as a distance between the mean values of the electron (in the excited state) and hole coordinates: $d^\ast\approx\langle z_e\rangle_{\rm exited} - \langle z_h\rangle$. For $F=0$, this value almost coincides  with the nominal center-to-center distance between QWs in the ADQW, $d=12.5$\,nm, but then it slightly increases with $F$, as the maxima of the electron and hole wave functions move in the opposite directions away from each other when the bias is applied. At larger EF, an anticrossing between electron ground  and excited states occurs at around $F=12.5$\,kV/cm, when both states have comparable probabilities for the electron to reside in each QW. Therefore, close to and after this anticrossing, the above equation for $d^\ast$ is no longer valid.

In order to evaluate precisely the e-h mean separation we take into account the contribution of both electron states into the IX (neglecting higher states, which can be also taken into account where necessary). To do so, we consider a $2\times  2$ matrix $z_{nm}$ defined by
\be
z_{nm}=\int \psi^e_n(z)\psi^e_m(z) z dz\,,
\tag{S\theequation}\label{znm}\stepcounter{equation}
\ee
where $\psi^e_n(z)$ is the wave function of $n$-th electron state, $n=1,2$ ($n=1$ for the ground and $n=2$ for the excited state), and the integration is done over the whole region of calculation of the wave function in $z$-direction (cutting off its exponentially growing tails [28]). Then we diagonalize the matrix $z_{nm}$ finding its eigenvalues $\langle z_e\rangle_{\rm min}$ and $\langle z_e\rangle_{\rm max}$ which play the role of the electron average position in the DX and IX, respectively, see Fig.\,\ref{fig:SM2}(d). The e-h mean separation in the IX is then given by
\be
d^\ast =\langle z_e\rangle_{\rm max} - \langle z_h\rangle\,.
\tag{S\theequation}\stepcounter{equation}
\ee
It is shown in Fig.\,\ref{fig:SM2}(d) demonstrating a weak monotonous dependence on the EF. These values of $d^\ast$ were used for determining the coefficient $\alpha$ standing at the exciton dipole moment ${\cal D}$
in Eq.\,(5) of the main text.

\section{Study of dipolaritons for other detunings}

\begin{figure}[b]
\hskip-0.0cm
\includegraphics*[angle=0,width=0.99\columnwidth]{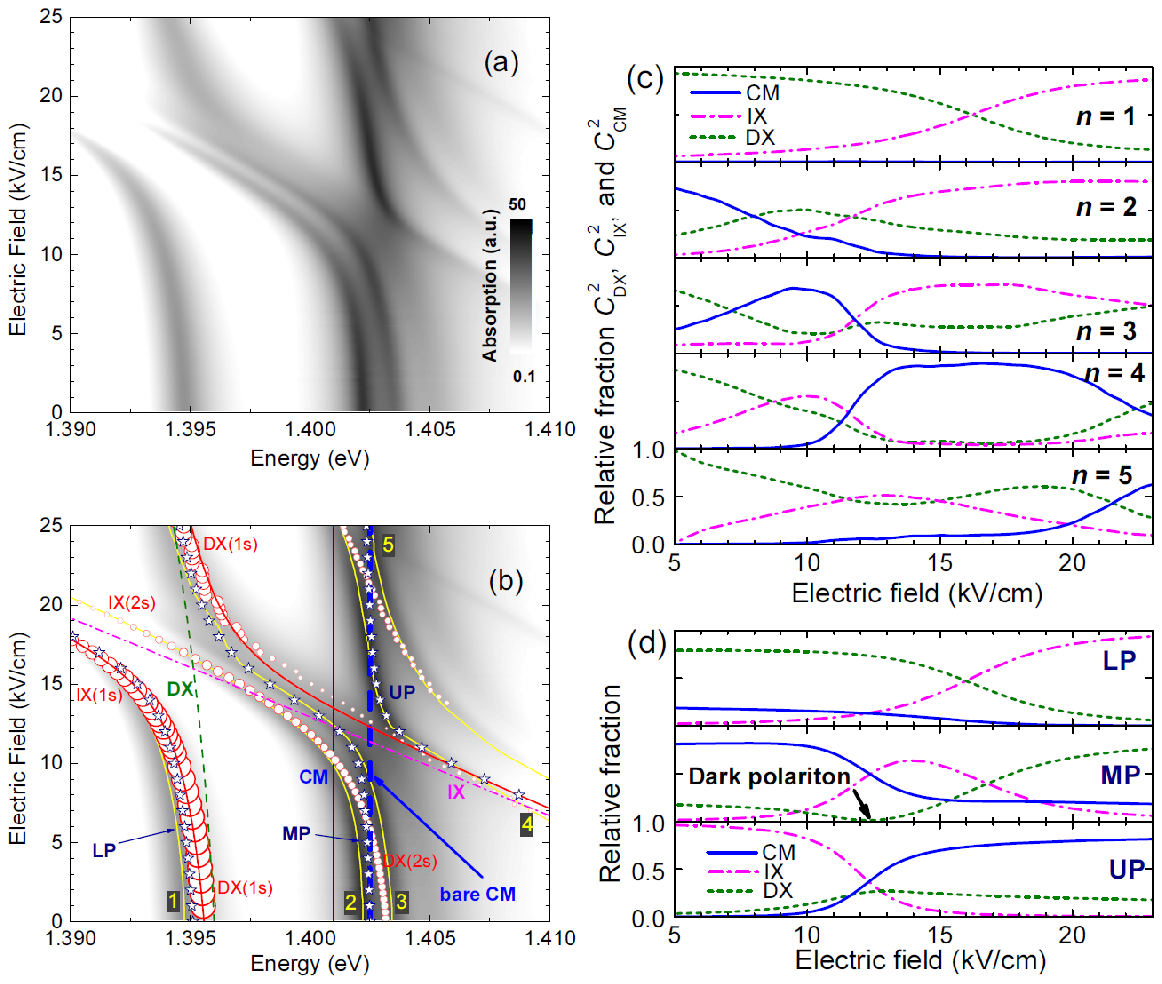}
\caption{
(a) Electric field dependent absorption spectrum of the microcavity-embedded ADQW structure on a logarithmic grey scale from white (0.1) to black (50),
for $\theta=35^\circ$ and the bare CM position at $E_{\rm CM}=1.4025$\,eV, i.e. for {\it positive} detuning. (b)-(d) The same as in Figs.\,4 and 5 of the main text but for $E_{\rm CM}=1.4025$\,eV. The absorption shown in (b) is exactly the same as in (a) where it is shown alone, i.e. without any additional data displayed in (b).
}\label{fig:SM3}
\end{figure}

\begin{figure}[b]
\hskip-0.0cm
\includegraphics*[angle=0,width=0.99\columnwidth]{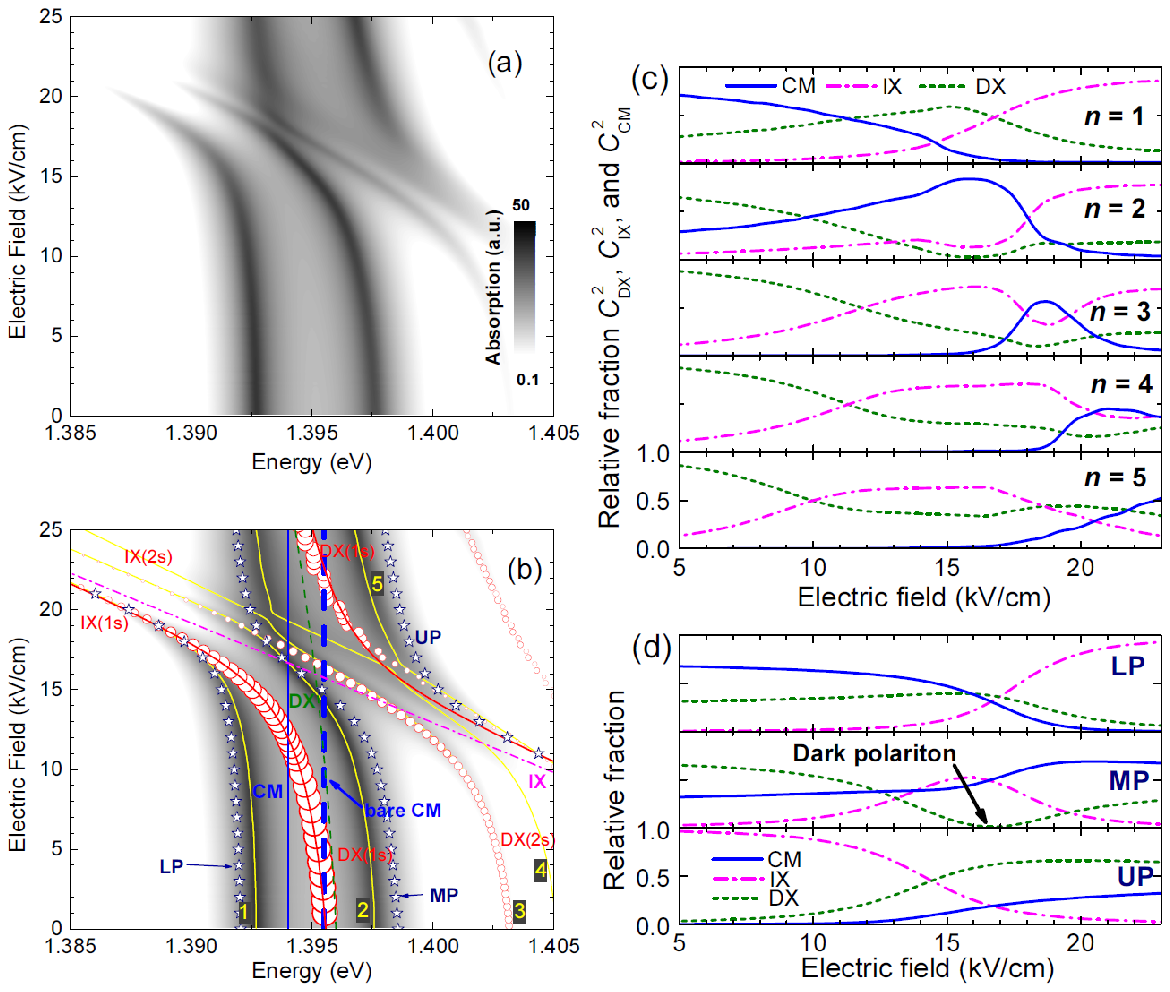}
\caption{ As in \Fig{fig:SM3} but for $E_{\rm CM}=1.3958$\,eV, i.e. for {\it zero} detuning.
}\label{fig:SM4}
\end{figure}

\begin{figure}[b]
\hskip-0.0cm
\includegraphics*[angle=0,width=0.99\columnwidth]{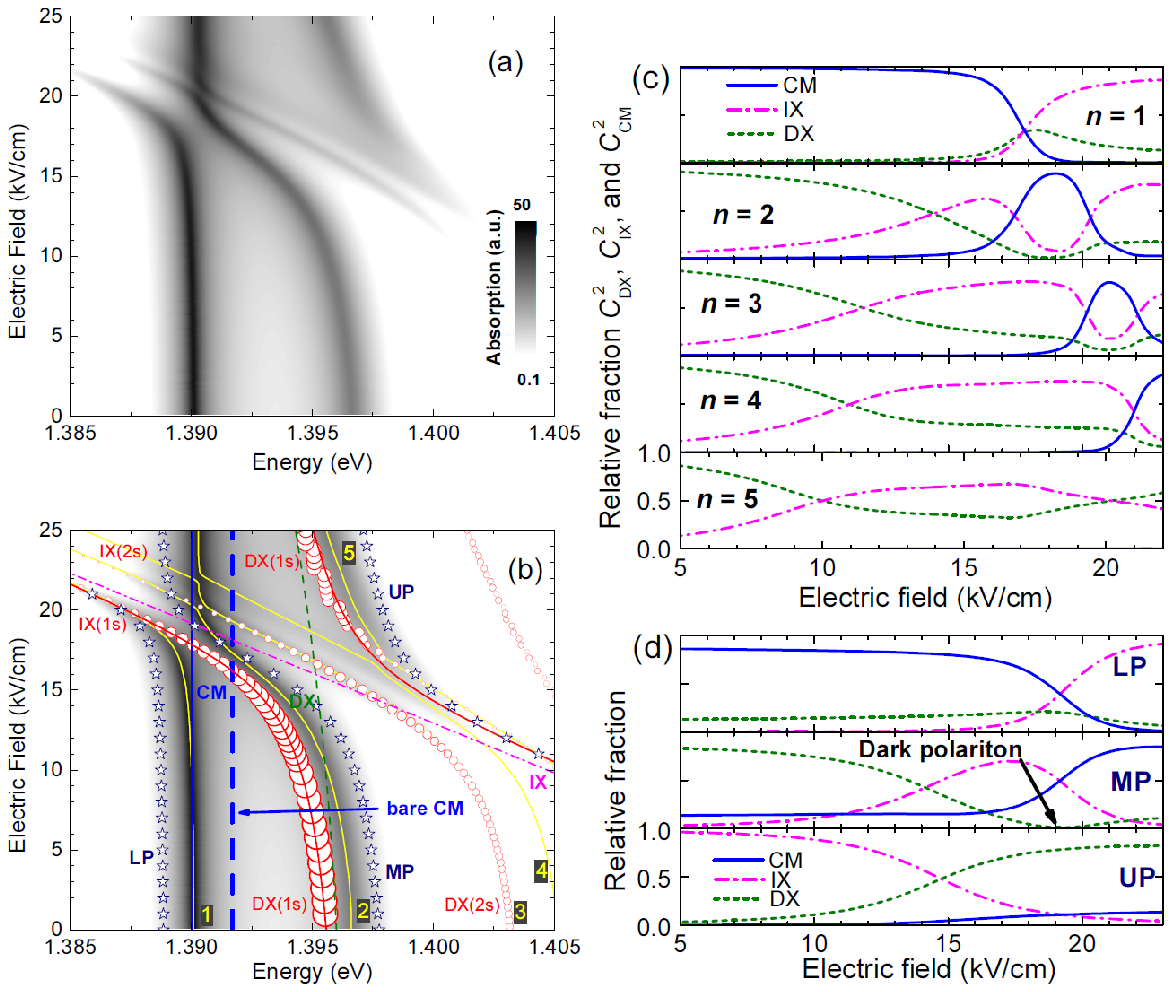}
\caption{  As in \Fig{fig:SM3} but for $E_{\rm CM}=1.3917$\,eV, i.e. for {\it negative} detuning.
}\label{fig:SM5}
\end{figure}

In this section, we show the analogues of Figs.\,4 and 5 of the main text for three different detunings: positive detuning giving the best fit to the experiment [20] (\Fig{fig:SM3}), zero detuning (\Fig{fig:SM4}) and negative detuning (\Fig{fig:SM5}). All three figures also show in panel (a) the absorption only,  with no extra lines on top of it, which is done for clarity of presentation. Within the TLM, the position of the dark polariton, shown in Figs.\,\ref{fig:SM3}-\ref{fig:SM5} (d) by an arrow, depends on the detuning, gradually shifting towards stronger EF when the detuning changing from positive to negative. This also demonstrates the fact that the dark polariton exists for any detuning, keeping zero contribution of DX and equal (50\%--50\%) contribution of IX and CM. This is in contrast with the microscopic calculation demonstrating clear changes in the properties of the dipolariton (dipolaritons are defined as polariton states with comparable contribution of IX and CM). Indeed, the DX fraction in the dipolariton reduces when the CM energy decreasing, compare the $n=2$ and $n=3$ polariton states for $F=10-12$\,kV/cm in \Fig{fig:SM3}(c) with those at $F=17-19$\,kV/cm in \Fig{fig:SM4}(c) and those at $F=19-21$\,kV/cm in \Fig{fig:SM5}(c). The DX fraction however never vanishes and never gets significantly smaller than the IX fraction when the latter remains comparable with the CM fraction, i.e. in a dipolariton.
Indeed, at the minumum of the DX component the ratio $C^2_{\rm CM}/C^2_{\rm IX}$ is usually maximized. For example, for zero detuning in \Fig{fig:SM4}(c), this minimum is observed at $F=15.8$\,kV/cm with $C^2_{\rm DX}=0.027$ and $C^2_{\rm IX}/C^2_{\rm DX}=4.8$, while $C^2_{\rm CM}/C^2_{\rm IX}= 6.4$ which is quite large.  Similarly, for negative detuning in \Fig{fig:SM5}(c), $C^2_{\rm DX}=0.017$ and $C^2_{\rm IX}/C^2_{\rm DX}=6.2$ at $F=18.6$\,kV/cm, but the IX component is again found to be very small: $C^2_{\rm CM}/C^2_{\rm IX}= 8.3$. All this clearly illustrates that the reduction of the DX component always results in the IX contribution also reduced as compared to the CM, so that such state can no longer be treated as a dipolariton.

Finally, we have found that the CM position obtained in the TLS fit is systematically (1.5\,meV) lower than the bare CM position, which is defined as position of the minimum of the reflection from the microcavity without the ADQW, compare blue solid and dashed vertical lines in Figs.\,\ref{fig:SM3}-\ref{fig:SM5}(b). This discrepancy is a clear indication of a significant influence of higher exciton states on the polariton properties and a weaker actual coupling between the DX and CM than predicted by the TLM.
Comparing panels (c) and (d) in Figs.\,\ref{fig:SM3}-\ref{fig:SM5} makes obvious the overall disagreement between the full microscopic calculation and the TLM.
All this demonstrates  the weakness of the TLM, in spite of the clear physical picture behind it and its success in prediction of the polariton energy levels.
\newpage

